\documentclass[final,5p,times,twocolumn]{elsarticle}

\usepackage{amssymb}

 \usepackage{lineno}
 \usepackage{color}
\usepackage{tikz}

\usepackage{subfigure}
\usepackage{textcomp}
\usepackage{gensymb}

\journal{Nuclear Instruments and Methods A}

\begin{document}

\begin{frontmatter}



\title{A Time Projection Chamber for High Accuracy and Precision Fission Cross Section Measurements}



\author[LLNL]{M.~Heffner\corref{cor1}}\address[LLNL]{Lawrence Livermore National Laboratory, Livermore, California 94550}
\author[PNNL]{D.M.~Asner}\address[PNNL]{Pacific Northwest National Laboratory, Richland, Washington 99354}
\author[CPOL]{R.G.~Baker}\address[CPOL]{California Polytechnic State University, San Luis Obispo, California 93407}
\author[INL]{J.~Baker\fnref{fn1}}\address[INL]{Idaho National Laboratory, Idaho Falls, Idaho 83415}
\author[OSU]{S.~Barrett}\address[OSU]{Oregon State University, Corvallis, Oregon 97331}
\author[OU]{C.~Brune}\address[OU]{Ohio University, Athens, Ohio 45701}
\author[CSM]{J.~Bundgaard}\address[CSM]{Colorado School of Mines, Golden, Colorado 80401}
\author[ISU,GIT]{E.~Burgett}\address[ISU]{Idaho State University, Pocatello, Idaho 83209}
\author[LLNL]{D.~Carter}
\author[LLNL]{M.~Cunningham}
\author[ISU]{J.~Deaven}
\author[CSM,LANL,CPOL]{D.L.~Duke}
\author[CSM]{U.~Greife}
\author[OU]{S.~Grimes}
\author[CSM]{U.~Hager}
\author[GIT]{N.~Hertel}\address[GIT]{Georgia Institue of Technology, Atlanta, Georgia 30332 }
\author[INL,LANL]{T.~Hill}
\author[ACU]{D.~Isenhower}\address[ACU]{Abilene Christian University, Abilene, Texas 79699}
\author[INL]{K.~Jewell}
\author[OSU]{J.~King}
\author[CPOL]{J.L.~Klay}
\author[ISU]{V.~Kleinrath}
\author[OU]{N.~Kornilov}
\author[CPOL]{R.~Kudo}
\author[LANL]{A.B.~Laptev}\address[LANL]{Los Alamos National Laboratory, Los Alamos, New Mexico 87545}
\author[OSU]{M.~Leonard}
\author[OSU]{W.~Loveland}
\author[OU]{T.~N.~Massey}
\author[ISU]{C.~McGrath}
\author[LANL]{R.~Meharchand}
\author[LANL]{L.~Montoya}
\author[ACU]{N.~Pickle}
\author[ACU]{H.~Qu}
\author[LLNL]{V.~Riot}
\author[LLNL]{J.~Ruz}
\author[LLNL]{S.~Sangiorgio}
\author[LLNL]{B.~Seilhan}
\author[ACU]{S.~Sharma}
\author[LLNL,CSM]{L.~Snyder}
\author[PNNL]{S.~Stave}
\author[PNNL]{G.~Tatishvili}
\author[ACU]{R.T.~Thornton}
\author[LANL]{F.~Tovesson}
\author[ACU]{D.~Towell}
\author[ACU]{R.S.~Towell}
\author[ACU]{S.~Watson}
\author[ISU]{B.~Wendt}
\author[PNNL]{L.~Wood}
\author[OSU]{L.~Yao}

\author{(NIFFTE Collaboration)}
\cortext[cor1]{Corresponding author, mheffner@llnl.gov}
\fntext[fn1]{deceased}

\begin{abstract}
The fission Time Projection Chamber (fissionTPC) is a compact (15~cm diameter) two-chamber MICROMEGAS TPC designed to make precision cross section measurements of neutron-induced fission.  
The actinide targets are placed on the central cathode and irradiated with a neutron beam that passes axially through the TPC inducing fission in the target.  
The 4$\pi$ acceptance for fission fragments and complete charged particle track reconstruction are powerful features of the fissionTPC which will be used to measure fission cross sections and examine the associated systematic errors.
This paper provides a detailed description of the design requirements, the design solutions, and the initial performance of the fissionTPC.
\end{abstract}


\end{frontmatter}


\section{Introduction}

Neutron-induced fission cross sections of the major actinides ($^{235}$U, $^{238}$U, $^{239}$Pu) have been studied for many years~\cite{White1967671,MEADOWSJW:FISCSP78,tovesson:pu239,NSR1974PO06,NSR1975CZ02,NSR2007NO07}. 
Evaluations of the cross sections are based on a large number of datasets and are thought to be very precise, better than 1\% in some cases, but the individual underlying datasets have uncertainties of 3--5\% in the fast neutron region (incident neutron energies from 100~keV to 14~MeV) and perhaps more significantly the individual experiments do not agree to within quoted experimental uncertainties~\cite{StaplesandMorley}.
The impact of cross section uncertainty has been studied extensively in the context of applications such as reactors, weapons and nucleosynthesis calculations~\cite{AlibertiG:Nucdsu} and it was concluded that uncertainties of 1\% or better are needed. 
In order to have confidence in the small uncertainties of the cross section evaluations and to understand the reasons for the spread in the current datasets, it is essential to perform a measurement with comparable uncertainty to the evaluation and that is as uncorrelated as possible to the previous measurements.
At the same time, the new experiment needs to be similar enough to previous experiments to explore the systematic errors of the previous experiments, and these sometimes competing needs have to be balanced.
The majority of fission cross section measurements have been conducted with fission chambers~\cite{fissionChamber}: simple, robust, easy-to-model detectors that have served the nuclear physics field well.
Although a list of possible error sources for the fission chamber experiments is well known~\cite{DesignStudy}, it appears further reduction of uncertainties is unlikely with traditional fission chambers~\cite{StaplesandMorley}.  

The NIFFTE (Neutron Induced Fission Fragment Tracking Experiment) collaboration has built a Time Projection Chamber (TPC)~\cite{nygrenTPC,TPCoverview}, the fissionTPC, to perform precision cross section measurements of the major actinides with the goal of better than one percent uncertainty.  
The focus of the fissionTPC design is to study uncertainties of previous measurements, including three of the largest sources of error: alpha particle and fission fragment differentiation, the target and beam non-uniformity, and the cross section uncertainty of the reference in ratio measurements (typically $^{235}$U).  
The unique experimental conditions also impact the design of the fissionTPC: the neutron beam passing through the detector, the high activity targets (of order MBq), and the large energy deposited by fission fragments.

\begin{figure}[htb]
  \begin{center}
    \includegraphics[width=\columnwidth]{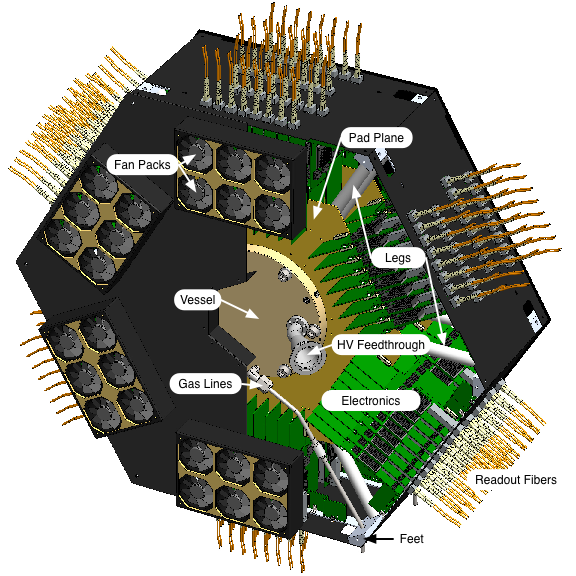}
    \caption{The fissionTPC with part of the fan pack and cover removed to see the inner components.}
    \label{fig:largeCutaway}
  \end{center}
\end{figure}

The fissionTPC is a MICROMEGAS (MICRO MEsh Gaseous Structure)~\cite{micromegas} TPC with 5952 hexagonal pads, 2~mm in pitch.  
A magnetic field is not required to control diffusion because of the short drift distance (54~mm).  
The argon-isobutane drift gas mixture is usually operated at 550~Torr where most particles of interest range out and stop in the active volume because of ionization energy loss.
Measuring this energy loss provides the total energy of each particle.    
Fig.~\ref{fig:largeCutaway} shows a cutaway drawing of the fissionTPC.  
The 15~cm diameter pressure vessel is supported at the center by 12 plastic legs that provide electrical isolation.  
All of the analog and digital processing is contained in the volume between the vessel and the end of the legs, an annulus approximately 16~cm wide (in radius) by 11~cm thick.  
Air cooling of the electronics is accomplished with fan packs.
Detector gas flows continuously through the fissionTPC via electrically isolated stainless steel tubes also supported by this structure.

\section{Design Requirements}

Ideally, the fissionTPC design will provide the capability to quantify the known and suspected systematic errors in previous fission chamber measurements (Table~\ref{tab_errors}) while maintaining the features that have been measured with low uncertainty in previous experiments (e.g. time of flight).  
To do this, the fissionTPC is designed to provide good 3D tracking of charged particles with near 100\% efficiency, and specific ionization measurements for particle identification.  
This section describes the design requirements related to addressing sources of systematic uncertainty as well as requirements related to the unique operational environment at the LANSCE neutron source. 

\begin{table}[htb]
\caption{Significant fission cross section systematic errors to be studied with the fissionTPC.}
\label{tab_errors}
\begin{center}
\begin{tabular}{l}
\hline
Sources of uncertainty\\
\hline
\hline
Particle ID, alpha/fission fragment separation \\
$^{235}$U reference \\
Energy loss in target\\
Neutron beam profile \\
Neutron beam energy-position correlations\\
Beam spreading and attenuation\\
Neutrons scattering back in (room return)\\
Target contamination \\
Non-uniform density (target and backing)\\
Complete fragment loss (detector efficiency)\\
\hline
\end{tabular}
\end{center}
\end{table}

\subsection{Particle Identification}

Perhaps the most significant contribution to the cross section uncertainty is the error in differentiating a fission event from a spontaneous alpha decay.  
In a fission chamber particles are identified based on the energy deposited in the chamber gas. 
One problem with this method is that the observed energy difference is narrowed by energy loss of fission fragments in the target.
This is a small effect for thin targets and for particles emitted perpendicular to the target surface.  
The energy loss for fragments emitted near-parallel to the target can be large enough that alpha particles and fission fragments are not distinguishable in a fission chamber; one has to rely on simulation to correct for this effect.  
In the case of $^{239}$Pu, the alpha decay rate is much larger than the neutron induced fission rate, increasing the possibility of misidentification.

Simulations indicate that measuring a nominal 20 ionization points along a typical track would be sufficient to identify the particles through differences in specific ionization, even when degenerate in energy.
The protons and alpha particles exhibit a Bragg peak at the end of the ionization track, while fission fragments produce the largest ionization at the start of the track.
This pronounced difference and the ability of the fissionTPC to measure the specific ionization is the key to particle identification.
High-resolution tracking also allows one to study the effects of energy loss in the target (the source of particle identification difficulty) as a function of emission angle, and make fiducial cuts to remove detector volume (e.g. shallow angles) that cannot be corrected.
Also important is sufficient dynamic range to measure the energy deposit of fission fragments and light recoil particles.
Preliminary calculations and measurements indicate that the 12-bit resolution is sufficient. 

\subsection{Target and Beam Uniformity}

A standard fission chamber does not have the ability to measure target uniformity.
The uniformity is measured outside of the fission chamber by examining variation in alpha decay rates across the surface.
The beam profile is measured with either neutron-sensitive film placed directly in the beam, or an external detector scanned through the beam.  
One challenge, and source of error, is the alignment of these different measurements.
An experimental simplification frequently used is to either make the beam larger than the target or the target larger than the beam, and assume that the larger item is uniform.  
This has the convenient result that one does not need to know the uniformity of the smaller item and the edge effects are removed; however the assumption of uniformity is questionable at the few percent level.  
In addition, the beam profile likely changes as a function of energy, which further complicates the matter. 
To improve on previous experiments, the fissionTPC will autoradiograph the target continuously, in situ, by tracking alpha particles from spontaneous decay.  
The beam profile is measured by monitoring the recoiling ions from neutron scattering on the drift gas: argon, carbon and hydrogen.
This allows one to measure both the target and beam uniformity with the same instrument at the same time and as a function of energy.  

The pointing resolution required to accurately characterize this uncertainty can be estimated by considering the expected variations in both the beam and target.  
Typical film exposures of a collimated spallation neutron beam show that the profile is smoothly varying and a resolution better than a few millimeters is sufficient.  
The target is expected to have thickness variations at nearly all length scales~\cite{target2} and is highly dependent on the method of depositing the material~\cite{Yaffe}.  
It is possible that large amplitude variations of target thickness occur on an area scale smaller than the fissionTPC pointing resolution.
If the amplitude of these variations is comparable to the range of fission fragments it could cause anomalies in the measurement.   
This effect cannot be corrected at all in a fission chamber. 
In a TPC with sub-micron resolution, the complication of small area, large thickness features vanishes because the range of the fission fragments in the target material is larger than a sub-micron feature.
It is not practical to build a TPC with sub-micron pointing accuracy but fortunately this is not necessary.
The expected pointing resolution from the fissionTPC is sufficient to split the target area into an ensemble of over a thousand patches that can each be analyzed and compared for consistency between the cross section and mass measurement for each patch; an observed inconsistency would indicate this pathology.
In addition to this comprehensive assessment of the entire target, it will also be characterized by scanning electron microscopy, atomic force microscopy or similar methods to supplement conclusions from the in situ measurements.

\subsection{Reference Target}

Most fission cross section measurements are relative measurements using a reference target along with the isotope of interest, so that the neutron beam flux, which is difficult to measure directly, partially cancels out of the calculation.  
$^{235}$U is a common reference, but the uncertainty from the evaluation (not including the issues with the evaluation discussed in the introduction) of the $^{235}$U cross section is close to 1\% at several energies between 150~keV and 20~MeV~\cite{Chadwick20112887}, making it impossible to measure a cross section of an isotope in ratio to $^{235}$U with a smaller uncertainty than one percent.
There are not many reactions that are known to better than 1\%, but the $^1$H(n,n')$^1$H reaction for fast neutrons below 14~MeV is known to 0.2\%~\cite{stoks_partial-wave_1993,carlson_international_2009}.  
TPC drift gases typically contain hydrogen (e.g. isobutane) and are very uniform in density which can be measured to the required accuracy.  
Plastic target backing materials could also serve as a redundant measure of the neutron flux with the same reaction.  
To make this measurement in ratio to $^1$H, the fissionTPC must be able to resolve protons and to measure the energy of the neutron that caused the proton recoil.  
The neutron energy will be measured in two ways, depending on the energy of the proton recoil: for low-energy protons that deposit their full energy in the gas, one can kinematically reconstruct the energy of the incident neutron using the proton energy and angular information; for high-energy protons that ionize the gas but don't stop in the chamber, neutron time of flight (TOF) will be used to reconstruct incident neutron energy, as is done for fission events (section~\ref{electronics}). 

\subsection{Other Sources of Uncertainty}

Outside of the three systematic uncertainties already discussed, there are a number of other potential systematic errors in existing data.
Full three-dimensional reconstruction of charged particles provides a powerful tool to study effects such as: neutron scattering effects, or room-return, which could be studied using targets placed inside the chamber but out of the direct beam; particles scattering from detector materials and edge effects; spallation; and (n,n') reactions.  
All of these effects can be studied in detail using the tracking capabilities of the fissionTPC.

\subsection{Operating Conditions}

A neutron beam passing through a TPC and fission occurring within a TPC are rarely considered in TPC design, but both are required for this application.
A number of design features related to these two operating conditions have to be optimized for a successful instrument: the beam size and time structure, the energy distribution of neutrons in the beam and the method to measure neutron energy, the effects of neutrons on the gain structures of the fissionTPC, the enormous ionization density of a single fission fragment, the average rate of ionization in the drift gas generated by alpha particles from a high specific activity target, and the 4$\pi$ emission of fragments.

The fissionTPC experiments take place on the 90~degree left flight path of target 4 at the Weapons Neutron Research (WNR) facility at the Los Alamos Neutron Science Center (LANSCE)~\cite{Lisowski2006910}.  
Although a TPC is capable of tracking many particles simultaneously and even of distinguishing overlapping events, a slow sweep of charge from the TPC (with respect to the beam spacing) complicates the analysis and increases space charge effects.
For these reasons, the fissionTPC requires a short drift time which is accomplished with a short drift length, although the fissionTPC must still be large enough to contain the full track for energy measurement.

The WNR neutron source generates neutrons via spallation, and provides a distribution of neutron energies within each proton pulse from the accelerator hitting the tungsten neutron production target.  
The energy for a given neutron is determined by its time of flight: the start signal comes from the accelerator signal, and the stop signal is observed by a reaction in the fissionTPC.  
The fissionTPC can be mounted between 6~to~8~meters from the spallation target which sets the required neutron timing resolution to be of order one nanosecond, which is not possible with the standard microsecond-scale TPC readout.
A fast cathode readout system is needed to measure the timing of a reaction in the fissionTPC to the required resolution.
The cathode readout serves also as the time start, or t$_{0}$, for the fissionTPC electron drift time measurement, which in typical high-energy TPC experiments is provided by an external trigger. 

The fission signals are very large, typically depositing 70--90~MeV (or about 10$^6$ electron/ion pairs in the gas) per fission fragment.  
This already provides a robust signal and the fissionTPC only needs a modest gain of 10--40 which is easy to obtain; however the Raether limit~\cite{raether_electron_1964,bay_study_2002} is easily exceeded and careful gas selection is needed to avoid sparking.
Although the fragment energy deposits are each very large, the rate is low (of order 10's Hz) and therefore they do not produce much integrated space charge.
The alpha particles from spontaneous decay are significantly less ionizing, but the rate (of order MBq) is high enough to produce significant space charge.
Space charge effects were studied with a simple model and determined to be modest for the amount of target material anticipated.  
The maximum deviation of charge drift due to space charge was estimated from simulated to be only 400--500~\textmu{}m for a 100~\textmu{}g/cm$^2$ $^{239}$Pu target assuming a MICROMEGAS ion leakage of 1\%.  
A laser calibration system has been designed to measure and correct for this effect.  

The fissionTPC is to first-order a 4$\pi$ detector with 100\% acceptance; the exception being the very narrow range of particles emitted parallel to the target losing most or all energy in the target.
The large acceptance is important for uncertainty quantification, as discussed previously, but it also has other advantages.
Since fission fragments are distributed nearly isotropically, the large acceptance increases the rate of data collection.
It also provides a complete picture of all charged particles associated with an event; it is easy to confirm a fission event with access to both fragments, ternary fission is clearly observed, pileup is easily distinguished and corrected, and any pathologies (e.g. spallation, bent tracks, distortions, etc...) are easy to identify.

\section{The fissionTPC Design}

\begin{figure*}[htb]
  \begin{center}
    \includegraphics[width=0.9\textwidth]{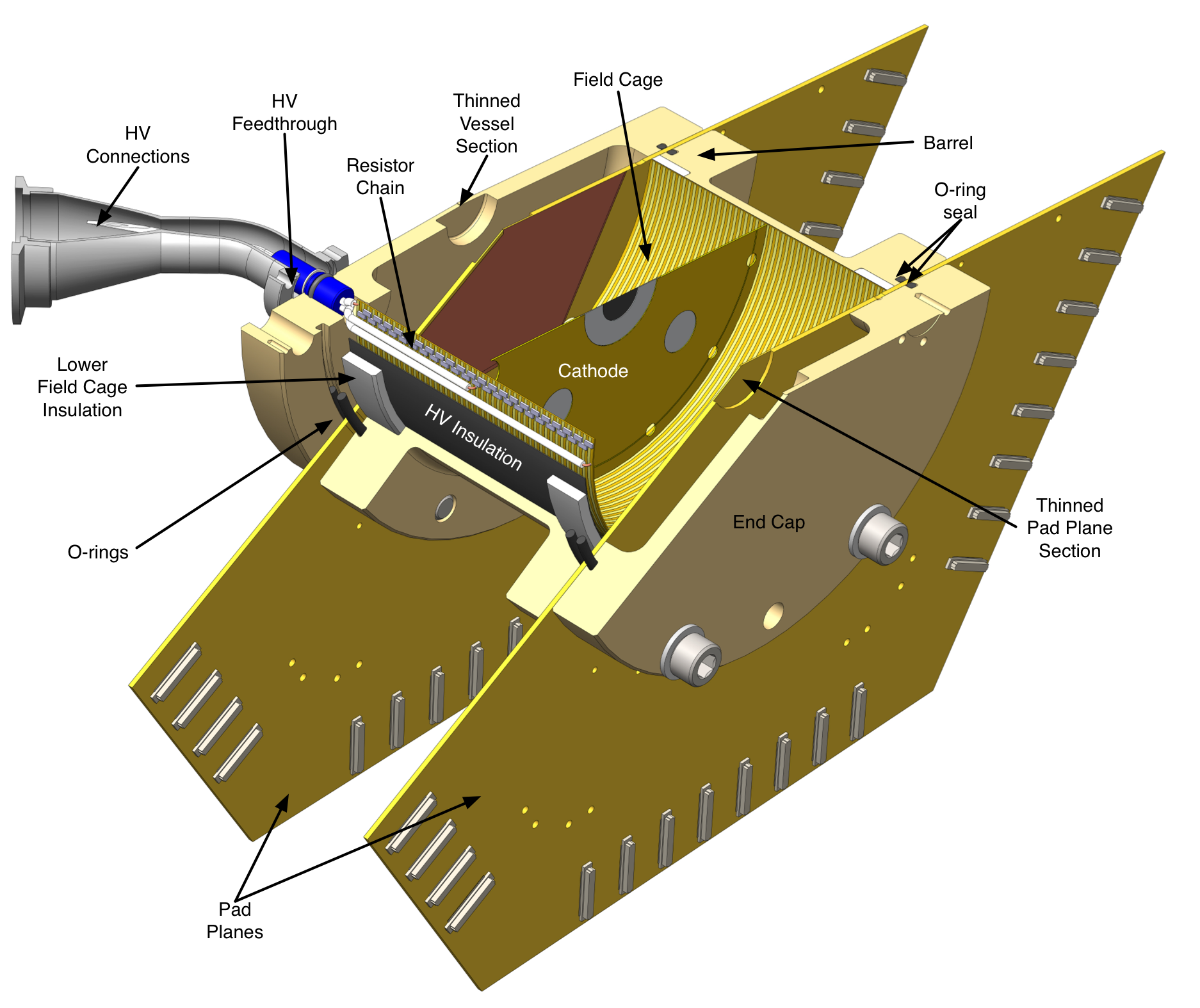}
    \caption{FissionTPC vessel half section showing the details of the chamber design.}
    \label{fig:vesselCutaway}
  \end{center}
\end{figure*}

A section view of the cylindrical fissionTPC chamber is shown in Fig.~\ref{fig:vesselCutaway}, and of some of the fissionTPC specifications are listed in Table~\ref{tab_parameters}.
The fissionTPC consists of two chambers surrounding the target to provide 4$\pi$ coverage.  
The cathode is common to each chamber, and is where the target is located.  
The pad planes and pressure vessel are thinned at the center to reduce the scattering of the neutron beam that passes through the center axis of the chamber.  
The pad planes are sealed with o-rings between the end-caps and the central body of the pressure vessel in a way that the center of the pad plane is in the drift gas and the edge is in the air outside of the fissionTPC.  
With this arrangement, the signals from the 5952 pads are transmitted through the inner layers of the pad plane providing a simple feedthrough for the large number of channels.

\begin{table}[htb]
\caption{Key design parameters of the fissionTPC}
\label{tab_parameters}
\begin{center}
\begin{tabular}{ll}
\hline
\textbf{Parameter} & \textbf{Value}\\
\hline
\hline
Number of Drift Chambers & 2 \\
Drift gases & H$_2$, P10, Ar/C$_4$H$_{10}$ \\
Gas pressure & 0--5~bar abs. \\
Magnetic field & none\\
Target diameter & 20~mm\\
Readout pads and pitch & hexagons, 2~mm\\
Nominal samples per track & 20\\
Nominal track length& 40~mm \\
Drift length& 54~mm \\
Number of pads & 5952\\
Drift voltage & up to 27~kV \\
Electron drift time & $\approx$1~\textmu{}s (P10) \\
\hline
\end{tabular}
\end{center}
\end{table}

The first design consideration is the orientation of the target and beam with respect to the fissionTPC.  
A cathode-mounted target is the most natural orientation for four primary reasons. 
Firstly, fission fragments and alpha particles from a target would be severely degraded passing through a window, so the target must be in the gas volume.  
Secondly, tracking distortions are difficult to avoid with the target outside of the TPC electric drift field, and the target must be on an equipotential to avoid distortions of the drift field.
Thirdly, fission chambers have a very similar geometry so a comparison of results (systematic errors in particular) between a TPC and fission chamber can be more directly made.
Fourthly, 4$\pi$ acceptance of fission fragments requires that the target be near the center of the TPC.
Besides the cathode-mounted target, a gaseous $^{239}$Pu target was also considered, as it would provide perhaps the best measurement possible, but most plutonium compounds are electronegative, which is a problem for gaseous ionization chambers.
Plutoniumacetylacetonate is a possibility, but it would require running the detector at elevated temperature ($\approx$200~\textcelsius{}).
The technical and operational risks associated with high-temperature gaseous-plutonium operation are high.  
      
With regards to beam orientation, the simplest method (and the one selected) is to orient the beam perpendicular to the cathode.
This has the added benefit of being most similar to a traditional fission chamber and therefore enabling the most direct comparison between the two techniques.
In this orientation the beam is parallel to the electric field (in a standard TPC configuration) which means the beam has to pass through the MICROMEGAS as well. 
Prototype tests indicate that this can be done without loss of MICROMEGAS function.
Other complicated options such as orienting the beam at an angle that misses the MICROMEGAS, or a radial drift field~\cite{ackermann_forward_2003} are not required. 
 
TPCs are generally slow readout devices due to the nature of the operation, namely, drifting electrons in gas.
This was a significant design consideration because the LANSCE accelerator operates at 1.8~\textmu{}s intervals between beam buckets.
Combined with the high alpha particle rate of some targets, this could produce excessive multiplicity and space charge accumulation.  
The readout time is determined by the drift gas, the electric field, and the physical length of the TPC.  
For any given gas and electric field, readout is always faster for a small TPC.  
Small TPCs have been built using a silicon chip readout~\cite{timepix} with a pitch of $\approx$ 55~\textmu{}m, but there is concern using this technology because the neutron beam would have to pass through the active part of the chips, causing bit upsets and latch-ups that could damage the chip in a short period of time.  
Somewhat larger micropattern readout structures such as the MICROMEGAS, the GEM (Gaseous Electron Multiplier)~\cite{sauli_gem_1997} and variants are inert structures without active components to be damaged by neutrons and are read out by a simple array of metal pads which would survive the neutron beam environment.  
Simple printed circuit board construction methods reliably produce 150~\textmu{}m features, therefore a 1--2~mm pad size is reasonable and also matches well with the beam spot of 20~mm and the range of particles in typical gases operating at typical pressures.  
After prototyping, a 2~mm hexagon pixel was selected.
The MICROMEGAS was selected over the GEM or LEM (Large Electron Multiplier~\cite{chechik_thick_2004}) for its simplicity of only one required voltage, low mass in the beam and low ion feedback~\cite{colas_ion_2004}.  

TPCs typically use a magnetic field to limit the electron diffusion and to bend the trajectory of particles to measure rigidity, which can be turned into a momentum once the charge state has been determined.  
The fissionTPC has a short 54~mm drift distance so the electron diffusion is already small and would not benefit greatly from a magnetic field.  
Since alpha particles and fission fragments stop within a few centimeters in a typical drift gas, the total energy is available by summing the ionization for a track.  
This is a far more direct measure than inferring the energy or momentum from the rigidity, which requires inferring the charge state.  
Due to energy loss and charge exchange from interactions with the gas, the rigidity is also continuously varying, complicating matters further and making a strong case for not using a magnet. 

\subsection{Pressure Vessel}

The pressure vessel was designed to accommodate pressures from vacuum to 5 bar absolute pressure for both inert and flammable gases.
The 6061T6 aluminum vessel consists of 3 parts: the central barrel that contains the field cage, and two end caps (Fig.~\ref{fig:vesselCutaway}).  
The gas lines and high voltage (HV) all pass through the end caps; no penetrations are made into the central barrel or the HV insulation, to preserve the HV integrity.  
The pad plane is sealed between the central barrel and end caps by the o-ring located in the flange.
This structure provides a gas-tight seal for the signal lines and the HV for the MICROMEGAS on each side.

To minimize beam scattering as it passes through the vessel, the center 25~mm diameter of each cap is thinned to 885~\textmu{}m for 5~bar operation (533~\textmu{}m for 2.5~bar operation).  
The gas is introduced through a 1/8''~NPT connection on one flange and is exhausted through an identical connection on the opposite flange.  
Additional NPT connections are used for pressure measurement and safety relief valves.  
The gas lines are fabricated from 304 stainless steel with welded VCR fittings and incorporate ceramic isolation to keep the vessel electrically isolated.

The space available for the HV feed-through to the cathode is limited and there exists no commercial connector that can satisfy both the density and HV isolation requirements, so a custom feed-through was developed (Fig.~\ref{fig:vesselCutaway}).  
A gas tight ferrule was constructed with an o-ring gland seal that seats into a 1~cm hole in the end cap and is held in place with a snap ring.  
Concentric with the hole is a threaded bolt pattern for a standard 1.33''~mini conflat that is used to bolt on a conduit that provides EMF shielding of the signals and a structural support for SHV connectors.  
Each flange has the 1~cm hole and mini conflat mount, the other port is used as the entry point for the calibration laser.

The vessel is supported by 6 pairs of 19~mm diameter Delrin tubes 26~cm long with electrically isolated all-thread compressing the tubes in place.  
The tube/all-thread supports connect the vessel to six aluminum feet which provide the support structure for the external electronics.
Two of the feet are used to mount the fissionTPC to a solid surface (Fig.~\ref{fig:largeCutaway}).

\subsection {Field Cage}

\begin{figure}
  \begin{center}
    \includegraphics[width=\columnwidth]{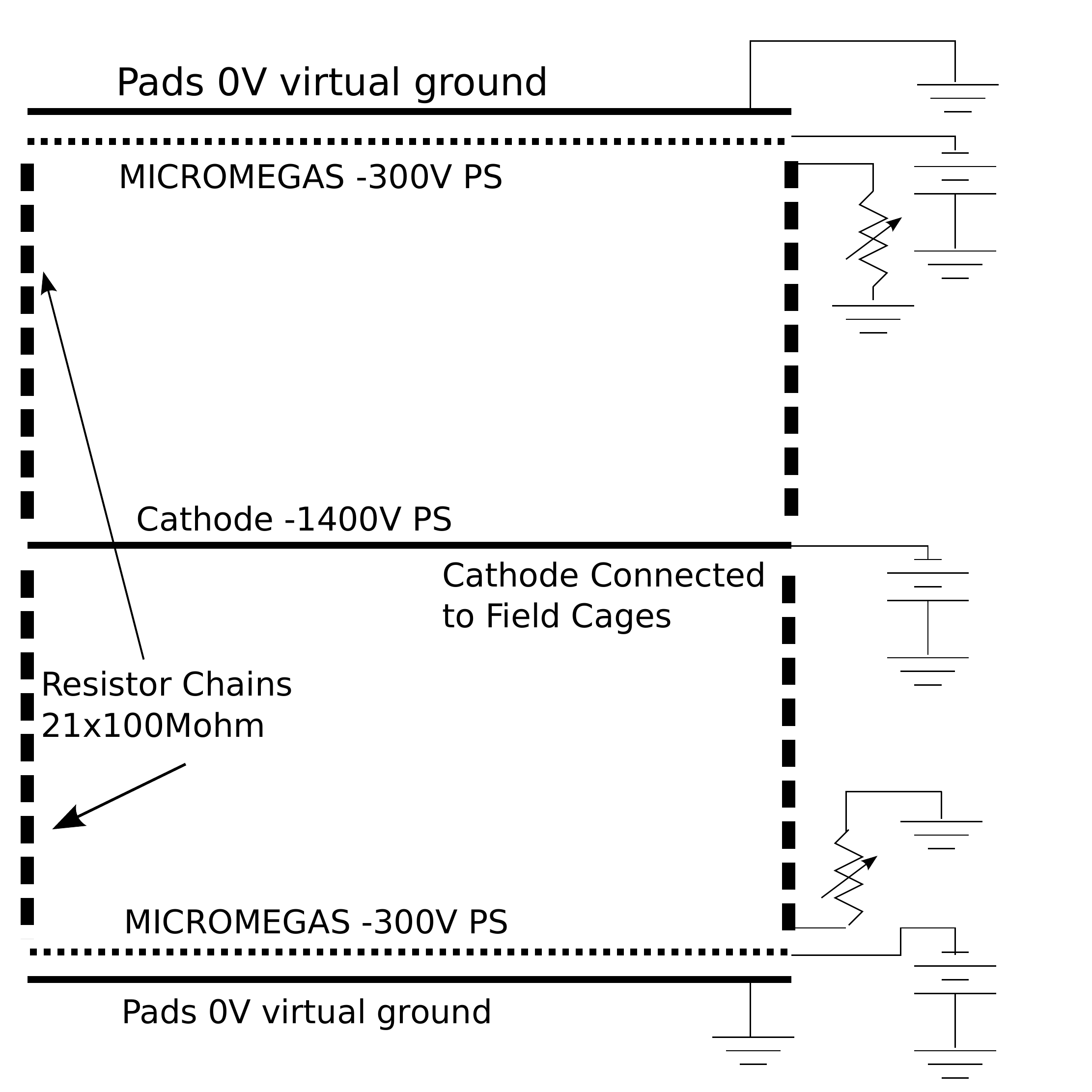}
    \caption{Biasing of the fissionTPC with nominal potentials shown.  
    Cross section view and dimensions are not to scale.  
    The adjustable resistors connected to the field cage are actually voltage controlled and the circuit is shown in Fig.~\ref{fig:fieldCageFollower}. 
    The field cage is set to be biased at the same voltage as the MICROMEGAS.  }
    \label{fig:fieldCageBias}
  \end{center}
\end{figure}

The field cage provides the shaping of the electric field to produce a uniform electric field over the active region of the detector.  
Fig.~\ref{fig:fieldCageBias} shows the biasing plan for the fissionTPC including the field cage.   
It is constructed from 1/2~oz/ft$^2$ copper clad G10 0.5~mm thick, commonly available from printed circuit board manufacturers.  
The 0.5~mm thickness provides good rigidity, but is still flexible enough that the flat material can be cut to size, clamped and epoxied (with low out-gassing TRA-CAST 3103) into a cylinder.

Stock PCB material is processed with standard techniques to produce copper lines that form rings of constant potential.  
The rings are then connected with a series of 100~M$\Omega$ resistors that step down the voltage (Fig.~\ref{fig:vesselCutaway}).  
The ring at the center connects to the cathode and the highest voltage is applied here.  
Each ring steps down the voltage on both sides until it reaches the micromesh voltage at the pad plane.  
Additional rings are placed on the outside of the field cage between the inside rings that step down in unison with the inside rings.  
This is a standard technique to provide good uniformity inside the active volume~\cite{blum}.

The field cage is 108~mm long (54~mm on each side) with an inside diameter (ID) of 144.14~mm and outside diameter (OD) of 145.14~mm.  
The difference between the OD of the field cage and the ID of the vessel (150~mm) allows room for insulation and electrical connections.
It has been tested in P10 and hydrogen at 5~bar up to 27~kV (needed because of the low mobility of electrons in hydrogen), although typical values needed for the argon-isobutane mixtures most often used are only a few thousand volts. 
The field cage is insulated from the pressure vessel with a 0.25~mm thick Cirlex sheet cut to the same size as the field cage, and the point where the field cage meets the pad plane is additionally insulated with a Teflon insert interference fit into the vessel (Fig.~\ref{fig:vesselCutaway}).

\subsection{Cathode}

\begin{figure}[htb]
  \begin{center}
    \includegraphics[width=\columnwidth]{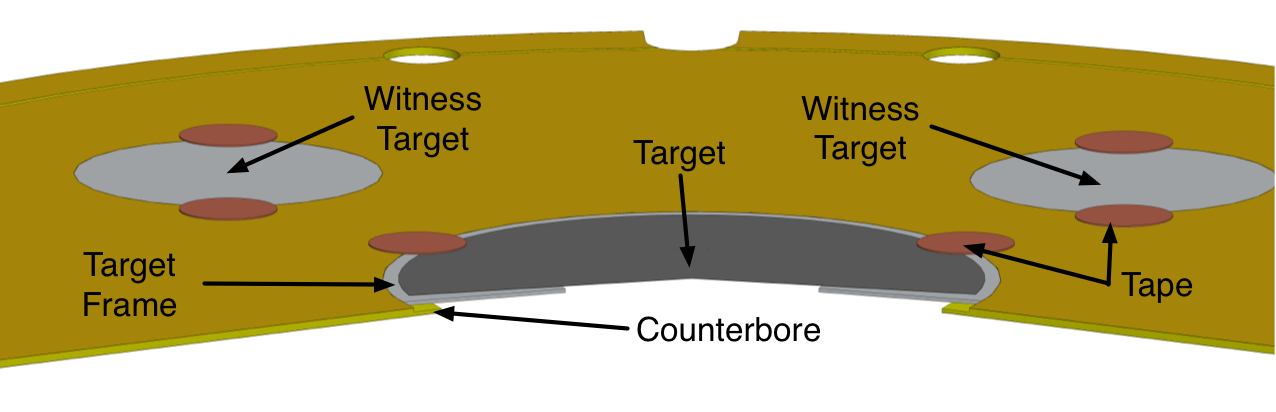}
    \caption{Cutaway of the fissionTPC cathode.}  
    \label{fig:cathode}
  \end{center}
\end{figure}

The conductive cathode is constructed from the same material as the field cage (Fig.~\ref{fig:cathode}).  
A 36~mm hole in the center of the cathode is designed to hold the target.  
The cathode hole has a 2~mm counterbore so the 40~mm target fits into the cathode and is held in place with copper tape dots 1/4'' in diameter (1181 series from 3M). 
The wires connecting the cathode and field cage to the HV feedthrough are routed between the field cage and vessel and cause the field cage center line to be shifted about 2~mm from the center line of the pressure vessel and beam. 
The location of the hole in the cathode is adjusted such that the target is on the axis of the pressure vessel and beam line.

In addition to the center target, four additional mounts for ``witness'' targets (20~mm diameter, or half the size of the center target) are located outside of the beam area and are loaded with $^{235}$U to measure the room scatter and other background effects in situ.
 
Both sides of the copper clad are connected with vias so it behaves as one conductor for both volumes, but has the required stiffness due to the fiberglass core.  
The copper is etched such that the outermost few millimeters are connected to the field cage, and the remainder is isolated with a 100~M$\Omega$ resistor and connected to an amplifier to read out a fast signal.  
The cathode is connected to a current preamplifier using a 1~nF 4~kV HV capacitor to bypass the HV and the preamplifier is located in the space behind the pad plane, but within the gas volume to keep the amplifier as close as possible to the cathode.
The short connection between the cathode and amplifier is important to limit the stray capacitance and inductance that interfere with the fast, low-noise response.

\subsection{Pad Plane and MICROMEGAS}

\begin{figure}[htb]
  \begin{center}
       \includegraphics[width=\columnwidth]{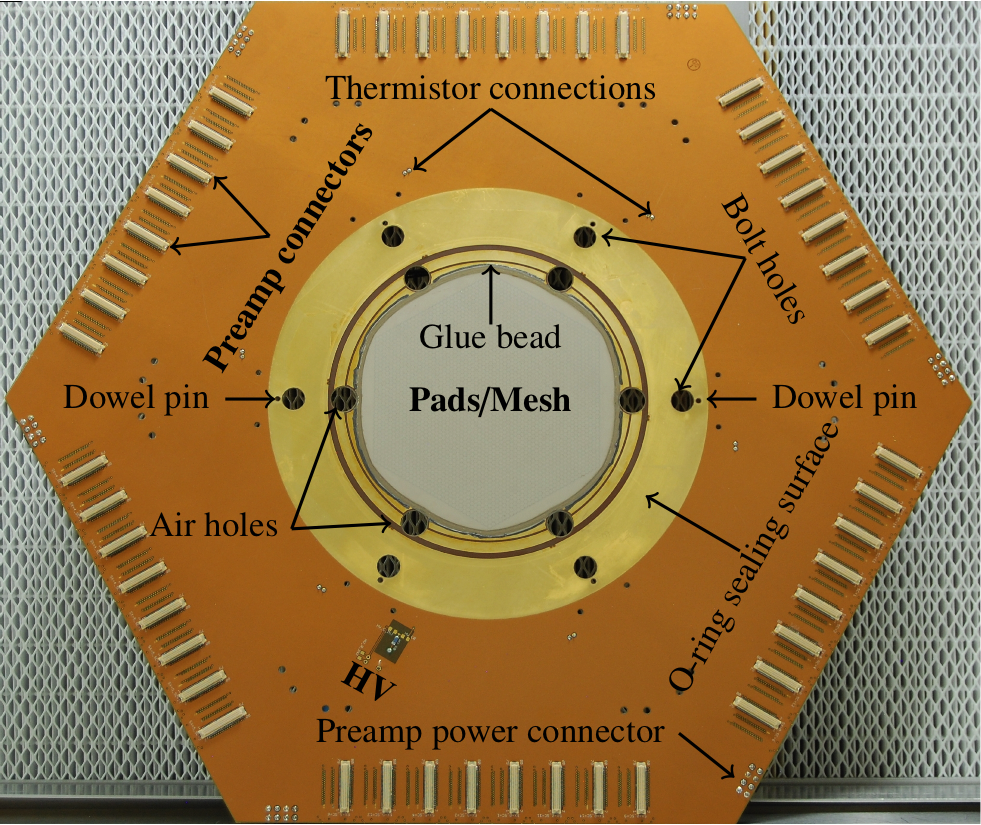};
    \caption{Overview of the pad plane. 
    The mesh covers the pads at the center and the glue bead that attaches the mesh to the pad plane is visible. 
	Thermistors measure the temperature of the pad plane in 6 places inside the gas volume. 
    The preamplifier connectors at the outer edge of the pad plane connect to the readout electronics. 
    For installation, two dowel pins are used to align with accuracy and, an o-ring surface on the pad plane assures the proper sealing of the drift chamber. 
    A collection of air holes release pressure differences between volumes. 
    }
    \label{fig:padPlane}
  \end{center}
\end{figure}

The pad plane is a 16 layer 1.57~mm thick printed circuit board (PCB) that is a hexagon 471~mm wide point to point (Fig.~\ref{fig:padPlane}).  
Each pad plane has 2976 close-packed hexagonal pads located at the center, support for the MICROMEGAS, and feedthroughs for additional signals such as thermistors, HV and preamplifer LV distribution.  
Each pad on the pad plane is individually connected to one pin on high density connectors at the edge of the board. 
The tracks from beam interactions originate from the center of the fissionTPC and an axially symmetric pad layout makes the best use of the expensive channels of electronics.

\begin{figure}
  \begin{center}
    \includegraphics[width=\columnwidth]{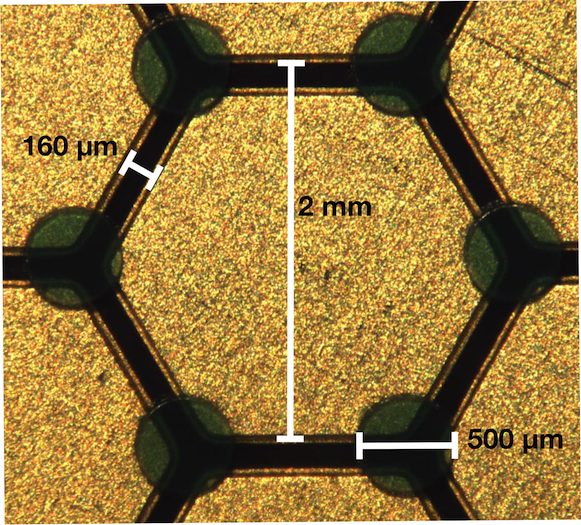}
    \caption{Micrograph of pad and pillar.  
    The gold colored hexagon at the center is one pad and the close pack tessellation of the neighbor pads with 160~\textmu{}m gaps between pads is also visible.  
    The green circle at each hexagon point is a 75~\textmu{}m tall insulating pillar to support the MICROMEGAS mesh.
    }
    \label{fig:pillarandPad}
  \end{center}
\end{figure}

A micrograph of a single pad is shown in Fig.~\ref{fig:pillarandPad}.  
The 2~mm hexagonal pad is formed by standard PCB photolithography and chemical etching of the 0.5~oz/ft$^2$ copper surface.  
Each pad has a 160~\textmu{}m via that is filled with copper and the pads are then ground to create a smooth flat surface.  
The board is then processed with immersion gold plating to passivate the surface and laminated with 4~mil of Vacrel 8100 series dry film solder mask and again photolithographically processed to produce the 500~\textmu{}m diameter, 75~\textmu{}m tall pillars at each point of each pad.  
Since beam passes through the pad plane it was thinned by removing 11 of the 16 layers with a counterbore at the center 2.5~cm.  
This reduces the thickness to the beam to approximately 500~\textmu{}m, but does not affect the pads on the surface.  
The board was designed using custom software to produce the Gerber files from which the board was manufactured by Streamline Circuits in Santa Clara, CA.

Since the insulating pillars are formed on the pad plane already by the PCB board manufacturer, to make a MICROMEGAS one only needs to lay a micromesh on the surface.  
The electroform nickel mesh is 3~\textmu{}m thick with 1000 lines per inch and is manufactured by Precision Eforming located in Cortland, NY. 
The large number of lines per inch was selected to minimize ion feedback~\cite{colas_ion_2004}.  
The low mass mesh is pulled down on the pillars by the electrostatic force from the normal operating voltages and therefore no additional processing is needed over the pads.  

The mesh is bonded to a conductor on the pad plane, outside of the pad area, that provides the HV to the mesh.
A channel is formed at the same time and of the same material as the pillars at the location that the mesh is to be bonded.
This acts as a dam to keep the epoxy from flowing away from the desired location.
TRA-BOND F113 epoxy, from TRA-CON, with 25\% graphite powder by weight is used to make the electrical connections between the mesh and the conducting surface on the pad plane.
This epoxy was chosen due to its low cost/quantity, low outgassing, low viscosity and conductivity without silver, a combination of traits that could not be found in pre-mixed epoxies. 
Silver was avoided because of its large neutron capture cross section and low viscosity was required to flow through the mesh when bonding it to the surface.  
Because the very thin mesh wrinkles easily, the manufacturer provides the mesh on a frame which can be laid on the pad plane and epoxied.  
When the epoxy cures, the frame is cut away.
All of the work is done on a clean bench that is approximately a class 100 clean area to prevent dust and other contaminants from interfering with the small 75~\textmu{}m gap.

The total capacitance of the 150~mm diameter mesh is calculated according to 
\begin{equation} \label{meshCapacitance}
  C = \epsilon \frac{A}{d} = 8.85 \times 10^{-6} \frac{\mathrm{nF}}{\mathrm{mm}} \frac{\pi (70\mathrm{mm})^2}{0.075\mathrm{mm}} = 1.8\mathrm{nF}
\end{equation}
where $A$ is the area of the mesh and $d$ is the height of the mesh above the pad plane.
The $\approx$ 2~nF capacitance ignores the effects of the gas, pillars and gaps between pads and was verified with an E4980A Agilent precision LCR meter.  
The HV is connected to this mesh via a 300~M$\Omega$ resistor, isolating the power supply in the event of a discharge.  
The energy of a spark is limited to the 2~nF discharge and preamplifiers are protected with a protection circuit that consists of a 65~$\Omega$ series resistor and a VBUS05L1-DD1 diode from Vishay Semiconductor.

The area of one pad ($\frac{\sqrt{3}}{2}W^2$ where $W$ is the width from flat to flat) is about 3.5~mm$^2$ resulting in a capacitance of about 0.4~pF per pad.  
Because the beam passes through the pad plane it is not possible to put the preamplifiers close to the pads, and some loss of signal occurs due to the capacitance of the trace from the pad to the preamplifer.  
Finite element analysis of the trace geometry sets this capacitance as high as 10~pF.  
This signal loss is offset by the large signals and the virtual ground of the charge sensitive amplifier.

The pad plane implements a simple but automatic addressing scheme so a readout channel can be assigned to a particular pad.  
This is done with 8 pins of the high density connectors; each connector is assigned an 8-bit number that is encoded by grounding the appropriate pins in the design of the PCB.  
This is read by the electronics and reported with the data.  
The advantage of this scheme is that the readout cards are interchangeable and replaceable without manual tracking of individual card locations.

\label{sec:pointingAccruacy}
The pointing and angular resolution can be estimated as described by~\cite{blum}.    
The pointing accuracy of a given track can be estimated from
\begin{equation} \label{pointingAccuracy}
\sigma_a = \frac{\epsilon}{\sqrt{N+1}} \left(\frac{12r^2+1+\frac{2}{N}}{1+\frac{2}{N}}\right)^{1/2},
\end{equation}
where $N$ is the number of points on the track, $r$ is a parameter related to the detector geometry (0.5 in this case).
Assuming a track with 20 points ($N$=20) and a single point accuracy of $\epsilon$=1~mm, the pointing resolution $\approx$~420~\textmu{}m.
The angular accuracy of the track can be estimated with 
\begin{equation}\label{angleAccuracy}
\sigma_b = \frac{\epsilon}{X_n}\frac{1}{\sqrt{N+1}}\sqrt{\frac{12N}{N+2}}.
\end{equation}
Incorporating an estimated track length ($X_n$) of 40~mm, an angular resolution of $\approx$~18~mrad is obtained.

\subsection{Drift Gas}

The drift gas serves as the medium in which the particles lose energy, ionization electrons drift, and gain in the MICROMEGAS is formed.  
It also serves as a target to measure the neutron beam with a light ion reaction like the well measured H(n,n)H.
Light gases such as hydrogen or helium are more sensitive to the neutron beam, due to the lower mass.  
Helium recoils are not ideal since they are identical to alpha particles emitted from the target and it is important to be able to distinguish the two.    
Hydrogen gas is not ideal because the electron drift is slow and therefore requires higher voltages.  
Hydrogen is also not well quenched and therefore has a tendency to break down.  
P10 (90\% argon, 10\% methane) is a cheap, fast, stable gas that can be premixed for ease of use and works well for high rates of highly ionizing particles.  
However, the MICROMEGAS sparked when a P10 filled fissionTPC was placed in the neutron beam.  
Therefore a 95\% argon, 5\% isobutane mixture that is more resistant to breakdown was used.  
A significant downside to the argon-isobutane mixture is that it can't be premixed at high pressure because of the low vapor pressure of the isobutane and will require instrumentation capable of accurately measuring the mixture if the hydrogen in the isobutane is to be used as a target.
This is yet to be designed for the $^{239}$Pu/$^1$H measurement.

\section{Target}

A number of targets are needed to fully characterize the systematic errors associated with this measurement.  
Actinide deposits of 50--200~\textmu{}g/cm$^2$ are supported on various backing materials such as thin carbon, plastic foils, and thick (0.25~mm) aluminum.  
To mimic the fission chamber ratio measurements (and the associated beam non-uniformity canceling), targets with a deposit on each side of a thick backing have been run where each side of the target is imaged in one chamber of the fissionTPC.
Since the tracking capabilities of the fissionTPC allow one to trace the origin of an event, multiple deposits have also been placed on the same backing. Thin backings allow for the coincident measurement of both fragments from a single fission event. 
A standard interface with the cathode was developed to make the target loading simpler and to accommodate the various target configurations.

With thin backed targets there is a significant risk of breaking, resulting in contamination of the fissionTPC that could be a health hazard and render the fissionTPC unusable.
The health hazard is addressed by loading and unloading the target from the fissionTPC in a glovebox and equipping the inlet and outlet gas lines with filters.
The minimum possible number of components have direct contact with the gas, so that in the event of a sample rupture, these components can be disposed of and replaced. The fissionTPC electronics, the most expensive part of the system, are located outside of the gas vessel and can be removed safely should a sample break inside. 

\subsection{Mechanical Design}

The standard target is a 0.25~mm thick Al disk that is 40~mm in diameter.  
Actinide deposits are made directly on one or both sides for ratio measurements.  
For some targets, a 20~mm diameter hole is made in the center of the Al disk where thin backing materials are mounted.  
So-called ``witness" targets consisting of 200~\textmu{}g/cm$^{2}$ deposits of $^{235}$U can be prepared on 0.25~mm Al disks (20~mm dia.).  

\subsection{Deposits}

Actinide deposits are made using either vacuum volatilization or molecular plating. 
Deposits of 100--200~\textmu{}g/cm$^{2}$ $^{235}$U and $^{238}$U were prepared using vacuum volatilization.
Starting with 99.99\% isotopically pure materials, tetrafluorides of the actinides were prepared by dissolving the oxides in HF, reducing them to the +4 oxidation state with stannous chloride and drying the resulting precipitate. 
The tetrafluoride is resistively heated in a tungsten boat in an evacuated chamber to produce the volatilization and masks were used to limit the deposit area.
The deposits are uniform ($\le$~1.5\% thickness variations across the deposit) and can be prepared on a variety of backing materials~\cite{target}.  
Atomic force microscopy of the evaporated targets shows rms surface roughness of 3~nm. 
In vacuum volatilization, the material being deposited is the tetrafluoride of the actinide element.

For rare or high specific activity isotopes, like $^{239}$Pu, molecular plating is used, due to its high deposition efficiency ($\ge$~90\%)~\cite{target2}.
Actinide nitrates in 4M HNO$_3$ are added to 15~ml isopropanol in a plating cell with the target as the anode and a Pd cathode.
The target is masked to produce the proper shape.
In molecular plating, the stoichiometry and crystal structure of the deposit is uncertain.  
Sadi et al.~\cite{target2} postulated a bridge structure with the overall stoichiometry of U$_{2}$O$_{6}$H$_{2}$ as consistent with their IR, Raman and x-ray spectroscopy of molecular plated U.

\section{Electronics and Readout}
\label{electronics}
The fissionTPC channel count (5952) combined with high data rates precludes the use of commercial electronics.  
A custom readout electronics solution was developed, called the EtherDAQ, and is detailed elsewhere~\cite{EtherDAQ}.  
The EtherDAQ contains the preamplifier and digital processing to provide a complete solution from pad to Ethernet packet. 
Each EtherDAQ card services 32 fissionTPC pads requiring a total of 192 cards for the entire fissionTPC.  
Each card requires about 0.5~A at 24~V for a total of about 100~A which is provided by an Agilent N8755A power supply.  
The preamplifiers on each EtherDAQ card require three voltages: 3.3~V (29~A), -3~V (61~A), and 7~V (54~A) which are provided by 5 Agilent 6651A power supplies.  
The total power consumed by the electronics is about 3.1~kW (2.4~kW digital, 650~W analog), and the required cooling is provided by a fan pack of 24 fans (Multicomp mc21692) and refrigerated air for high altitude operation at LANL.
The cards are organized into banks of 16 cards and the power is distributed through a custom distribution chassis.  
This chassis also distributes the common clock, trigger, sync pulse, and system trigger holdoff.  
The power and signals from the distribution chassis are further distributed to the cards on the pad plane for the preamplifiers and by a small bus board for the 24~V digital part.

The neutron time of flight from the cathode signal and accelerator needs to be measured with timing resolutions of order 1~ns, and needs to be synchronized with the main clock of the fissionTPC.  
This was done by an EtherDAQ card with a special preamp card that sends sequentially delayed signals to 20 channels of the card.  
Each channel receives a signal delayed 1~ns from the previous channel using the Data Delay Devices 1520SA-100-500 chip, effectively providing a GHz sampler with the same timebase as the main fissionTPC clock.  

The amplifier used to read out the cathode must be fast for time of flight measurement, but also low noise in order to measure proton recoils without gas amplification.  
A current amplifier, a modified CR-100 from Cremat in Newton, MA, is used instead of a charge amplifier due to the high rate.    
The modification is the replacement of the 100~M$\Omega$ feedback resistor with a 50~k$\Omega$ resistor.  
This compact amplifier fits within the fissionTPC volume, and can amplify alpha particle and fission fragment signals with reasonable performance.   
A custom amplifier is under development to achieve better sensitivity and reduce pickup from the accelerator. 

Each EtherDAQ card is connected via a 1~GB fiber to Dell 6224F Ethernet switches, which are linked together and connected to the data aquisition (DAQ) computer through a single 10~GB uplink.  
A single desktop computer (8~core Xeon x5550 with 6~GB memory) processes the data from the cards and stores it locally.

The fissionTPC can operate in two triggering modes: a global triggered mode, where a trigger causes all channels to readout; or a triggerless mode where each channel reads out when that channel exceeds a threshold.
The latter is used for most operations.
The alpha decay rate from the $^{239}$Pu target is about 5 orders of magnitude larger than the fission rate of interest.  
To suppress the alphas, a trigger hold-off is used to limit data collection to periods when beam is present.  
This suppresses $\approx$~97\% of the data volume , leaving data rates of 20--30~MB/s for sources with MBq activities.

\subsection{DAQ Software} 

The EtherDAQ cards implement a simple data transfer method based on the raw Ethernet protocol.  
The data collected by the EtherDAQ is stored in an array of memory locations on the EtherDAQ.  
A state machine on the FPGA continuously transmits each packet in turn and continues until a confirmation packet is sent by the receiving computer.  
Once the acknowledge is received the memory location is freed for the next data packet.  
The receiving computer has a packet receiver that receives the data and sends the acknowledge packets.  
The receiving software also repackages the data into sequential time order and displays summary statistics to determine the quality of the data.

\subsection{Data Management}

The typical data volume for a 3 week run with a MBq source is $\approx$~50~TB, and there are 4--5 such runs per year.  
The data is stored locally in an AberNAS 50~TB RAID and is continuously transferred via internet to an off-site high performance storage system (HPSS).  
Attached to the HPSS is a 1152 core (12.9~TFLOP) computer cluster that processes the data into summary data sets for analysis.
Slow data (temperatures, voltages, etc..) are written to a PostgreSQL database along with run information that is then used in the data processing. 

\subsection{Offline software}

The software paradigm adopted for this project is based on compartmentalizing the TPC-specific processing components into a separate library (which will later become the basis for an open-source TPC software project) and the fissionTPC-specific components that implement a software framework for this experiment.  

Before selecting tools for implementing the framework, a review of modern standards in high energy physics software was undertaken.  
This involved studying the software and computing models used by the PHENIX and STAR experiments at RHIC~\cite{PHENIX:Software,STAR:Software}, LHCb and ALICE at the CERN LHC~\cite{LHCb:Software,ALICE:Software}, E907/MIPP at Fermilab~\cite{E907:Software} and BaBar at SLAC~\cite{BaBar:Software}. 
Almost all of these experiments use the C++ language and the CERN ROOT data analysis framework~\cite{CERN-ROOT} for data storage (persistency), histogramming and analysis.
The exception to this rule is LHCb, which chose to separate the memory-resident or transient data from the persistent data with no explicit dependence on ROOT software libraries.
Based on this review, the fissionTPC project adopted a hybrid software model written in C++ with internal transient data objects and ROOT-based persistency.
The transient data classes defined in the TPC library each have a corresponding persistent data class with ROOT streamer information in order to be read/written to ROOT files.

At present, track finding and fitting are accomplished with modules that implement the Hough Transform~\cite{Hough} and Kalman Filter~\cite{Kalman}.  
Since there is no magnetic field, the track model is a simple straight line fit. 
However, other algorithms such as 3-D edge finding followed by 3-D least squares minimization, and 2-D cluster/hit finding and ``follow-your-nose'' tracking are also available.
All of the modules implementing these algorithms are interchangeable, allowing one to use any of the track-finding modules with any of the track-fitting algorithms.
Having multiple reconstruction algorithms is essential for estimating systematic uncertainties on the final output, especially critical for the high precision results expected from the fissionTPC.

The fissionTPC event display is based on TEve, ROOT's OpenGL visualization tool, and TGeo, ROOT's 3D geometric modeling package.
The level of detail available in the event display can be adjusted by the user at the command line. 
Images can also be zoomed, rotated, panned, scanned and otherwise manipulated with the mouse.  
Beyond visualization of the detector geometry, the event display provides the capability to visualize data events and allows the user to interact with and interrogate the visual information.  
Some example images from the fissionTPC event display are included in section~\ref{sec:performance}.

Further details about the software framework, calibration, reconstruction and analysis algorithms will be included in future publications on fissionTPC data.

\section{Auxiliary Systems}

In order to operate and monitor the fissionTPC a number of auxiliary systems are required.

\subsection{Gas System}

The fissionTPC was designed to use a number of gases including flammable gases such as hydrogen and isobutane.
All of the gases used are inexpensive and non-poisonous so a once through system that vents the gases to the atmosphere was selected.
The gas consumption rates for the small ($\approx$2~liter) fissionTPC volume are modest (less than 100~cc/min in operation) and standard $\approx$200~ft$^3$ cylinders provide more than enough gas for a standard 3 week run.
Premixed gases are used when practical (e.g. P10), but in the case of isobutane the low vapor pressure makes this difficult and it is mixed with other gases using the fissionTPC gas system.

The gas system uses MKS 1479A mass flow controllers on the input and output lines and a 627B pressure gauge to maintain pressures and flow rates in the fissionTPC.
The pressure is monitored with a high accuracy 120AA Baratron pressure gauge.
It is controlled by MKS 647C and PR4000 controllers using the internal proportional-integral-derivative (PID) controller to set and maintain the pressure and flow rate. 
The controllers are connected to a Fedora 12 Linux computer to provide remote control of the controllers and to record the conditions. 

The gas is filtered at the inlet and outlet of the fissionTPC with Mott filters (GSG-V4-1-N) that provide 0.003~\textmu{}m filtering while maintaining a pressure drop of only 0.1~psi at the nominal flow rate of 0.1~liters/min.
After the last flow controller of the gas system the exhaust goes through a vacuum pump that allows the system to operate sightly below atmospheric pressure.
The gas then passes through a gas quality monitor, a health physics filter, and vents outside.

The gas quality monitor measures trace water and oxygen content.  
The water is measured with a Pura Hygrometer from Kahn Instruments and typically reaches $-60$~\textcelsius{} to $-70$~\textcelsius{} dew point (10 to 3~ppm) during operation of the fissionTPC.  
The oxygen is measured with the OpTech Oxygen Sensor from MoCon.  
This fluorescence quenching device is designed to measure headspace oxygen in food packaging.  
It does not need to be replaced at regular intervals and can withstand atmospheric oxygen levels without damage, unlike common electrochemical sensors.  
The disadvantage to using this sensor is a loss in sensitivity around 100~ppm, and therefore during operation the sensor reports the minimum oxygen measurement.  
The short electron drift time in the fissionTPC places a modest requirement on electron lifetime so 100~ppm oxygen levels are sufficient for the fissionTPC operation and measurement below this value is not necessary.    
The sensor was placed in a clear section of tubing and the readout instrument placed outside.
Automated control and readout are accomplished with a Raspberry Pi single board computer.

\subsection{Calibration Systems}


Radon gas was introduced into the fissionTPC as a method of channel gain correction.  
The radon decays uniformly throughout the fissionTPC volume and the resulting time-averaged pulse responses from the amplifiers are corrected to match, since the input signal is the same on all channels.  
The radon ($^{220}$Rn) source was made from natural thorium ore crushed up and placed in an inline filter that prevents any dust from entering the fissionTPC, but allows the passage of radon gas out of the filter.  
The filter is placed inline to the inlet gas supply.  
The half life of the radon is short (55~s) so the gas line has to be kept short and a reasonably high gas flow maintained but this also assures that the radon is gone from the chamber quickly when the source is turned off.  
This technique is similar to the $^{83m}$Kr used in other TPCs~\cite{afanasiev_na49_1999}; the advantage of radon is that the alpha decay energies are closer to the energy scale required for this fissionTPC calibration.
Preliminary operation of this system, with a sample of about 500 channels, shows a total channel to channel variation (including amplifiers and structural variations in the MICROMEGAS) of around $\pm$10-15\%.

To calibrate electric field non-uniformities and effects of space charge, a 266~nm 4~mJ laser system was designed to provide tracks in the fissionTPC at specific locations to sample distortions.
The interface on the fissionTPC and some of the hardware has been prepared, but the system has not yet been deployed.

\subsection{Slow Controls}

This section describes the instruments used to control the experiment and record slow data.
All components are connected to Ethernet and controlled through the MIDAS~\cite{midas} interface.
MIDAS provides the web interface to start and stop data collection, power up/down the detector, and provides plots that track slow data to monitor the experiment. 
This configuration also allows the fissionTPC to be operated and monitored offsite.
Some of the slow data was originally recorded with MIDAS, but that proved to be cumbersome and writing the data directly to a database offers a cleaner solution.
The slow data can then be viewed from the database with a custom web interface.

\subsubsection{High Voltage Bias}

\begin{figure}
  \begin{center}
    \includegraphics[width=0.8\columnwidth]{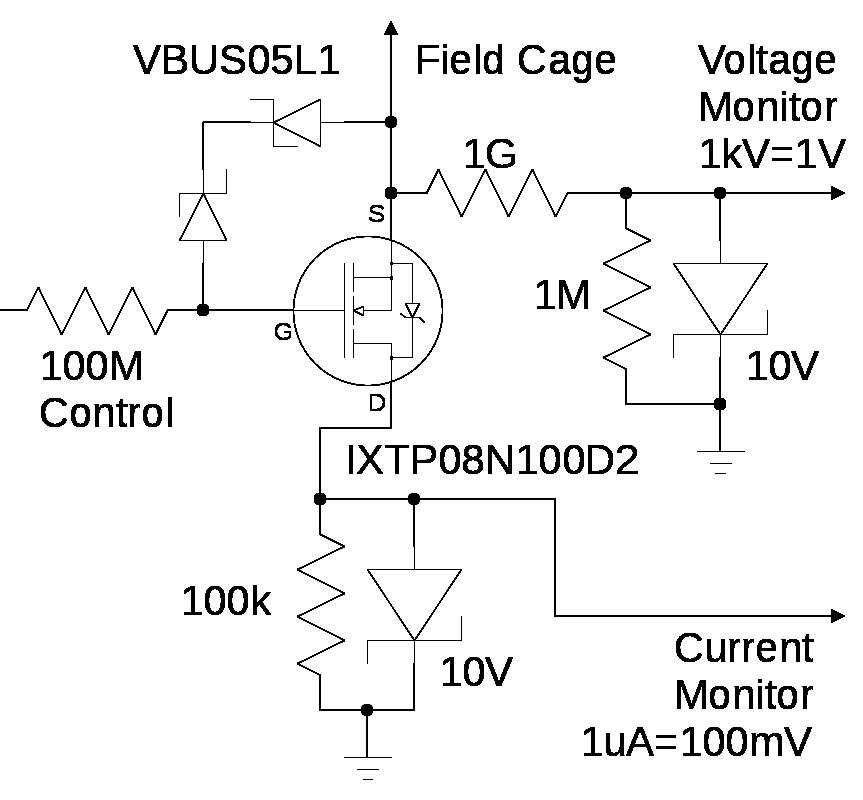}
    \caption{Circuit that sets the potential of the anode plane of the field cage.}
    \label{fig:fieldCageFollower}
  \end{center}
\end{figure}

The high voltage is supplied with four SRS PS350 5~kV power supplies and one SRS PS370 20~kV power supply connected to a Prologix GPIB-to-Ethernet adapter.
The SRS PS370 supplies the cathode and the four SRS PS350 supply each MICROMEGAS and each field cage voltage follower. 
This configuration allows complete remote control of the biasing of the fissionTPC and automated ramping of the voltages.
The noise level of the SRS supplies induces a measurable signal when connected without filtering.
The cathode supply is filtered with a 20~M$\Omega$ resistor and 0.5~nF capacitor in a low pass filter configuration.
The cathode voltage and current are measured from the digital interfaces to the SRS supply.
The ground path of the field cages is connected with the circuit shown in Fig.~\ref{fig:fieldCageFollower}.
This circuit also provides a means to measure the current through the field cage and the voltage at the low potential end (anode).
The MICROMEGAS is connected with an RC filter consisting of 300~M$\Omega$ resistor and the capacitance of the MICROMEGAS which is about 1.8~nF.
The 300~M$\Omega$ resistor limits the current in the event of a spark, but the power supply does not have a low enough trip current.
A custom circuit was constructed to measure the current, which is recorded continuously, and to trip the supply in the event of a spark.

The anode of the field cage is not at ground potential, but is held at the same voltage as the micromesh for the MICROMEGAS.  
It is not electrically connected to the MICROMEGAS to prevent the field cage current from inducing noise on the pads, but it must have the same potential to keep the field lines straight.  
The anode of the field cage needs a path for the current and a method to adjust the current.  
Most power supplies are designed to source a current and have limited ability to sink a current, so a simple custom circuit was designed (Fig.~\ref{fig:fieldCageFollower}). 
This is a simple FET follower circuit using a 1~kV IXTP08N100D2, and is controlled by a voltage from a power supply and the effective resistance of the FET is set automatically to place the anode voltage at the control voltage of the circuit.  
The voltage on the MICROMEGAS rarely needs to exceed 450~V, so the 1~kV rating is sufficient, but there are other devices that can go to larger voltages.

\subsubsection{Low Voltage Power}

The kilowatt-scale power delivered to the electronics is monitored by a custom power and clock distribution unit (PCDU).
This box is controlled remotely by a Keithley 2701 controller that can shut down the power to the fissionTPC in the event that monitored temperatures exceed safety thresholds or by the command of remote monitoring shift personnel.

\subsubsection{Temperature}

The pad planes temperatures are measured in up to 12 locations inside the gas volume with GE Thermometrics DC95F103W thermistors connected to a Fluke 1560 Black Stack controller.
The Fluke 1560 Black Stack controller was selected because it has low enough excitation current that the themistors in proximity to pads do not induce a measurable cross talk signal on the fissionTPC pads.
Temperatures are also measured at various locations outside of the fissionTPC gas volume using a Keithley 2701. 

\section{Performance}
\label{sec:performance}
A number of prototypes were constructed and studied between 2008 and 2012 to arrive at the final design.
The full fissionTPC was completed early in 2013 and initially operated with sealed sources, followed by operations in-beam starting August 2013.
This section is a summary of the results of the prototyping and early performance results from the fissionTPC.
Cross section measurements and other physics results will be the subject of subsequent publications. 

\begin{figure}[h]
  \begin{center}
    \includegraphics[width=\columnwidth]{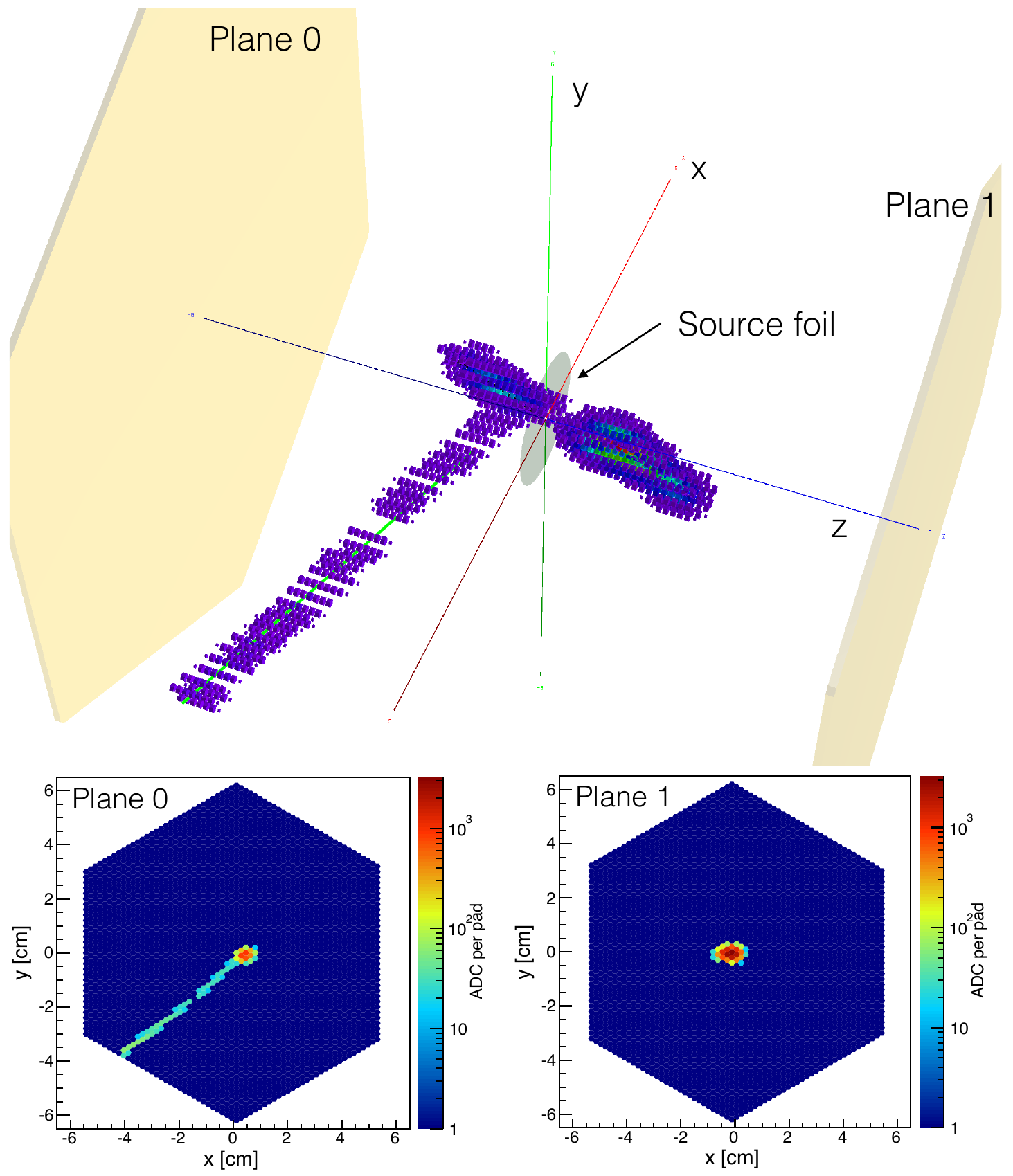}
    \caption{An alpha-accompanied spontaneous fission event from $^{252}$Cf in the fissionTPC.  
    The color of the voxel indicates the amount of charge: blue is sparse charge and red is dense charge.
    The upper plot shows the fission event.
    The lower plots show the projection of the same event on each pad plane.
  }
    \label{fig:tracks}
  \end{center}
\end{figure}

The power of the fissionTPC is derived from the 3D imaging and specific ionization measurement capabilities. 
Fig.~\ref{fig:tracks} is an image generated from fissionTPC data of a single event which demonstrates some of this capability.
The recorded charge for each voxel is displayed in 3D.
Clearly, many tracks can be differentiated simultaneously, many pathologies can be spotted easily, and the particle can be identified by the ionization profile.  
This example is just one of millions of similar events that have been recorded. 
There are large numbers of events that yield the expected protons, alpha particles, and fission fragments, but there are also other interesting events such as particles hitting the edge of the target holder, (n,n') reactions on the argon and carbon in the gas, spallation from high energy neutrons, and events that would constitute pileup in a fission chamber but are clearly distinguished in the fissionTPC.

\subsection{Signals}

The primary signals from the fissionTPC are read out from the charge-sensitive preamplifiers attached to each pad.  
Because the beam passes through the center of the pad plane it is difficult to locate the preamplifers close to the pads.
In the worst case, the distance between the pad and the input FET of the amplifier is about 17~cm.  
The rise time of the amplifiers is not significantly degraded by the extra trace capacitance.
The typical rise time is dominated by the diffusion of charge, resulting in rise times of order 100~ns.  
The 12-bit ADC typically only reports a few counts of noise (peak-to-peak) and the threshold of most channels is set around 5 counts.  
This delivers the large dynamic range required to measure protons and fission fragments.  
The cross talk between the electronics is less than 0.5\%.

\subsection{Gas Gain and Energy Resolution}

\begin{figure}
  \begin{center}
    \includegraphics[width=\columnwidth]{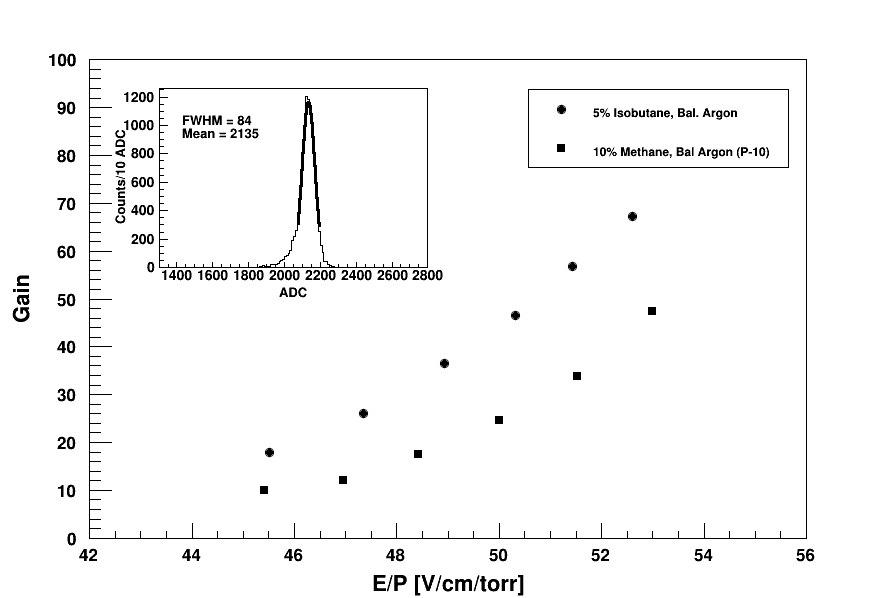}
    \caption{The measured gain vs reduced electric field for two gas mixtures, 95\% argon 5\% isobutane at 580~Torr and P10 at 880~Torr.  
    	Inset shows the alpha peak and fitted width from which the gain and energy resolution are determined.
    	A large component of this particular resolution measurement is known to be uncorrected gain variations of each pad.  
    	}
    \label{fig:alphaPeak}
  \end{center}
\end{figure}

A typical fission fragment ionizes a few million atoms in the drift gas. 
The electrons from the primary ionization drift to the anode where diffusion and track geometry typically spread the electrons out over a few 10's of pads.
A single pad could receive on order of 10$^5$ electrons before gas gain.  
The gain produced by the MICROMEGAS increases the number of electrons at the readout pad and has to be carefully selected to avoid breakdown in the MICROMEGAS while also allowing the amplification of signals from the less ionizing particles (e.g. protons).
The Raether limit is around 10$^7$--10$^8$, so a gain of just 100 produces a very large signal that in some cases could produce a breakdown.
This is the upper limit for operation with a fissioning source in the chamber and is confirmed by the observation that the same MICROMEGAS without a fissioning source will go to much higher voltages than one with a fissioning source.
The lower limit is set by the noise of the amplifier combined with capacitive losses transporting the signal to the amplifier, diffusion and the energy deposit of the lesser ionizing particles.
The lower limit is also an indicator of the dynamic range required for the readout, which the electronics has to accommodate.
In practice the gain is adjusted until the smallest feature of interest is well above noise at a gain of about 30.
The measured gain curve and alpha peak are shown in Fig.~\ref{fig:alphaPeak}.

The energy resolution is not a critical parameter for the cross section measurement, but it is a measure of how well the system is operating and may be of interest for other measurements such as A and Z determination of the fragments.
The resolution in a low Fano factor gas such as argon ($\approx$0.2) is expected to be dominated by the gas gain which has a form similar to the Fano equation but has an $f$ value closer to 0.7 in the case of a large charge swarm~\cite{gain_variance}.
An estimate of the FWHM energy resolution for a fission fragment after gas gain, where $N_e$ is the number of electrons and F is the Fano factor, is  

\begin{equation}
\label{gain_variation}
\frac{\delta E}{E} = 2.35\sqrt{\frac{F+f}{N_e}} = 2.35\sqrt{\frac{.2+.7}{10^6}} = 0.22\% .
\end{equation}

This is quite small and will therefore be dominated by the electronics, channel-to-channel variations, and the energy straggling in the target material.

In order to measure energy resolution, the voxels of ionization from an alpha track have to be summed over all of the readout pads that recorded charge from the particle.
The pad-to-pad total gain variations (including amplifiers, and structural variation of the MICROMEGAS) play a significant role in deteriorating the measured energy resolution and have been measured to be about 10--15\%.  
This variation can be corrected, but because this is not a priority for the cross section measurements it has not been done at this time.  
The inset of Fig.~\ref{fig:alphaPeak} shows alpha particle energies from which the gain (from the peak location) and the energy resolution of 4\% FWHM (from the width) have been determined.

\subsection{Neutron beam profile measurement}


The neutron beam profile is measured by looking at reactions in the drift gas because the uniformity of the gas is much better than a thin solid target.  
The gas is not restricted in the volume so there is virtually no pressure variance across the volume and the temperature difference across the fissionTPC is less than about 1.5~\textcelsius{} in the worst case which could produce only about 0.5\% density variation.  

\begin{figure}
  \begin{center}
    \includegraphics[width=\columnwidth]{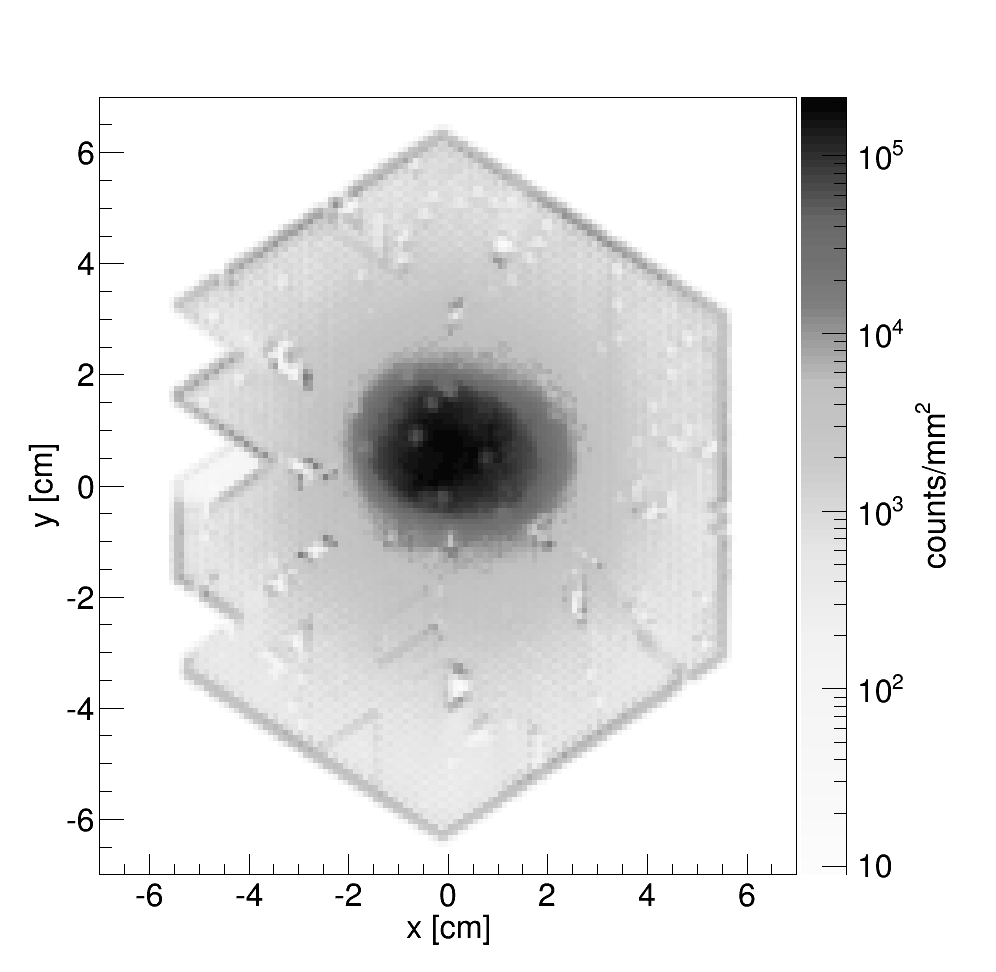}
    \caption{The beam profile measured by plotting the track start points which are dominated by the (n,p), C(n,n')C, and Ar(n,n')Ar reactions.  
    A few sectors are missing on the left, resulting in the jagged edge.
     }
    \label{fig:beamProfile}
  \end{center}
\end{figure}

There are two categories of reactions that can be used for this measurement.
Because of the neutron energy distribution of the beam, (n,n') reactions on the carbon and argon in the gas mostly produce very short tracks.
The charged particle recoils from these reactions only travel a few mm in the gas and therefore can not be resolved as tracks, but the charge deposited is large enough that they can be clearly identified and the location of the charge is a measure of where a neutron in the beam was located. 
In another category are tracks that are long enough to be measured (e.g. (n,p)) and in these reactions the start of the track is a measure of the neutron location.  
Using both of these reactions a plot is generated of the beam spot integrated over energy (Fig.~\ref{fig:beamProfile}).  
The beam collimators used for this beam profile measurement are round and have 20~mm diameter bores. 
The resulting beam profile is not round and the long axis of the profile is tilted from the horizontal.
This can be understood by considering the source of the neutrons is not a point source and the collimation is acting like a pin-hole camera for the neutrons.
The spot is an image of the spallation target.

Previous measurements of the beam profile with film indicated that the profile would be smoothly varying with only large scale variation at the cm scale and this is what is measured with the fissionTPC as well.
The fissionTPC pointing resolution sets the smallest scale that can be investigated for features and appears to be sufficient to fully characterize the beam profile for the calculation of the cross section.  
An additional feature of the fissionTPC is the ability to also study the beam profile as a function of energy using (n,p) reactions.
The neutron energy will be calculated from the proton energy and angle combined with the known kinematics of the reaction.
For protons that do not range out in the gas, the cathode readout will be used as a time of flight indicator.
The beam profile could change with energy and this method allows for compensation of this effect.

\subsection{Autoradiograph measurements}

\begin{figure}
  \begin{center}
  \subfigure[Fission fragment vertices]{
    \includegraphics[width=0.47\columnwidth]{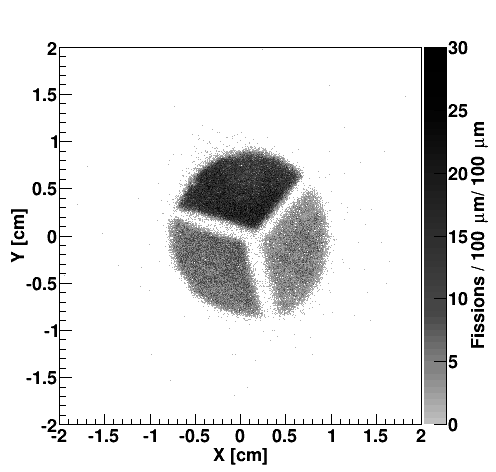}
	\label{fig:FFvert}
	}
\subfigure[Alpha particle vertices]{
    \includegraphics[width=0.47\columnwidth]{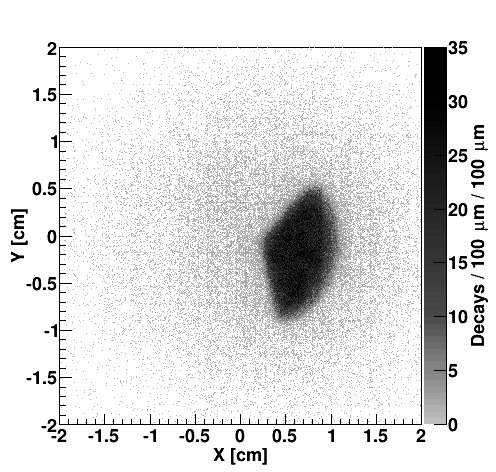}
	\label{fig:alphaVert}
	}
    \caption{Reconstruction of the actinide deposits of the same target by plotting the track starting vertex of fission fragments~(\ref{fig:FFvert}) and alpha particles~(\ref{fig:alphaVert})  The target consists of three wedges each of a different actinide, $^{238}$U, $^{235}$U and $^{239}$Pu.  The high spontaneous decay rate of the $^{239}$Pu is evident in \ref{fig:alphaVert}.}
        \label{fig:autoradiograph}
  \end{center}
\end{figure}

The target thickness is typically determined by looking at the alpha decay as a function of the location on the target, known as an autoradiograph.  
The fissionTPC makes an autoradiograph in situ simultaneously with beam profile measurement; an example is shown in Fig.~\ref{fig:autoradiograph}.  
Similar to the beam profile, the autoradiograph is generated by plotting the track start point, but in this case using alpha particles.
The ultimate pointing resolution is expected to be a few hundred microns.  

\subsection{Particle Identification}

To date particle identification has focused on differentiation of alpha particles from fission fragments, as is necessary for ratio cross section measurements.  
There are also efforts to unambiguously identify protons as well as the A and Z of the fission fragments.  
The identification is accomplished by quantifying the differences in the specific ionization of the various particles. Ultimately, each track will be compared with a $\chi^2$ test to ionization models and the probability of the identity of each track will be calculated.

\begin{figure}
  \begin{center}
    \includegraphics[width=\columnwidth]{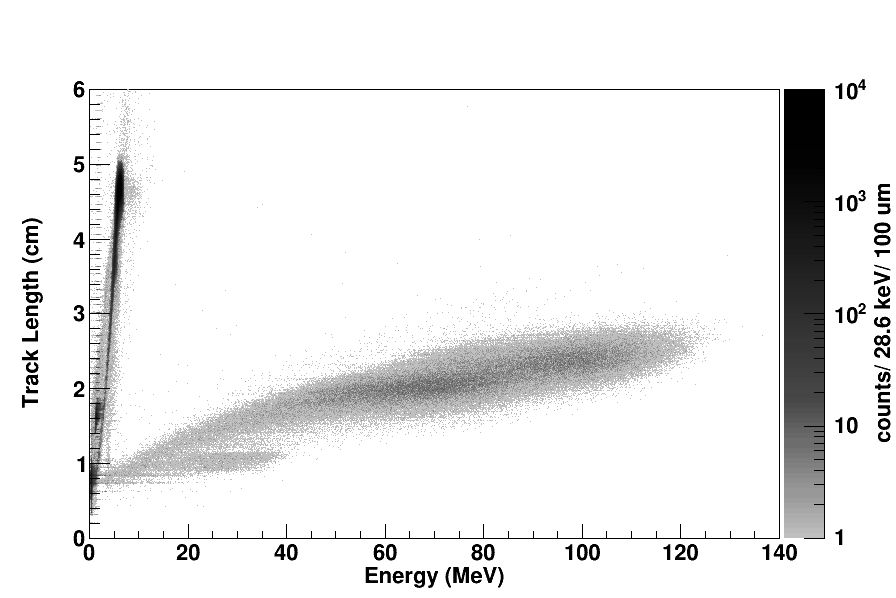}
    \caption{The length vs energy from a $^{252}$Cf source.  
    The alpha particles form a nearly vertical band and the fission fragments are nearly horizontal.
      }
    \label{fig:lengthEnergy}
  \end{center}
\end{figure}

Before a full model fitting analysis is complete, a simple particle identification can be accomplished by looking at the length and energy of a track.
Fig.~\ref{fig:lengthEnergy} shows a typical result from the fissionTPC using a $^{252}$Cf source. 
There are a number of interesting features in this plot, but there are two dominant features related to alpha particle and fission fragment separation.
The large horizontal band with track length around 2~cm and energies from 40 to 120~MeV are fission fragments.
The heavy and light fission fragments show up as regions of higher intensity centered around 70~MeV and 100~MeV.
The thin mostly vertical band at and below 6~MeV is the alpha particles.
In both cases, the bands extend down to zero energy due to energy loss of the particles in the target.
Tracks that exit nearly parallel to the target (and therefore traverse more material) exit the target with little kinetic energy. 
In the case of a fission chamber one would only have the total energy information, but in the displayed figure the additional separation of the two particle types is provided by the different energy deposit per track length.
 
Taking a closer look there is an additional band around 1~cm with energies from 0 to 40~MeV, and a dot at 1.7~cm with an energy of about 3~MeV.
Close investigation of these tracks indicate that they are particles that hit the edge of the target holder, an edge only 250~\textmu{}m tall.

The same plot has also been generated for in-beam data, and the same features persist.
The in-beam plot is more complicated because of the increase in the number of types of particles one can also identify: protons, alpha particles at higher energies than the decay alpha particles, and (n,n') interactions.

\subsection{Fast Cathode Measurements}
 
\begin{figure}
  \begin{center}
    \includegraphics[width=\columnwidth]{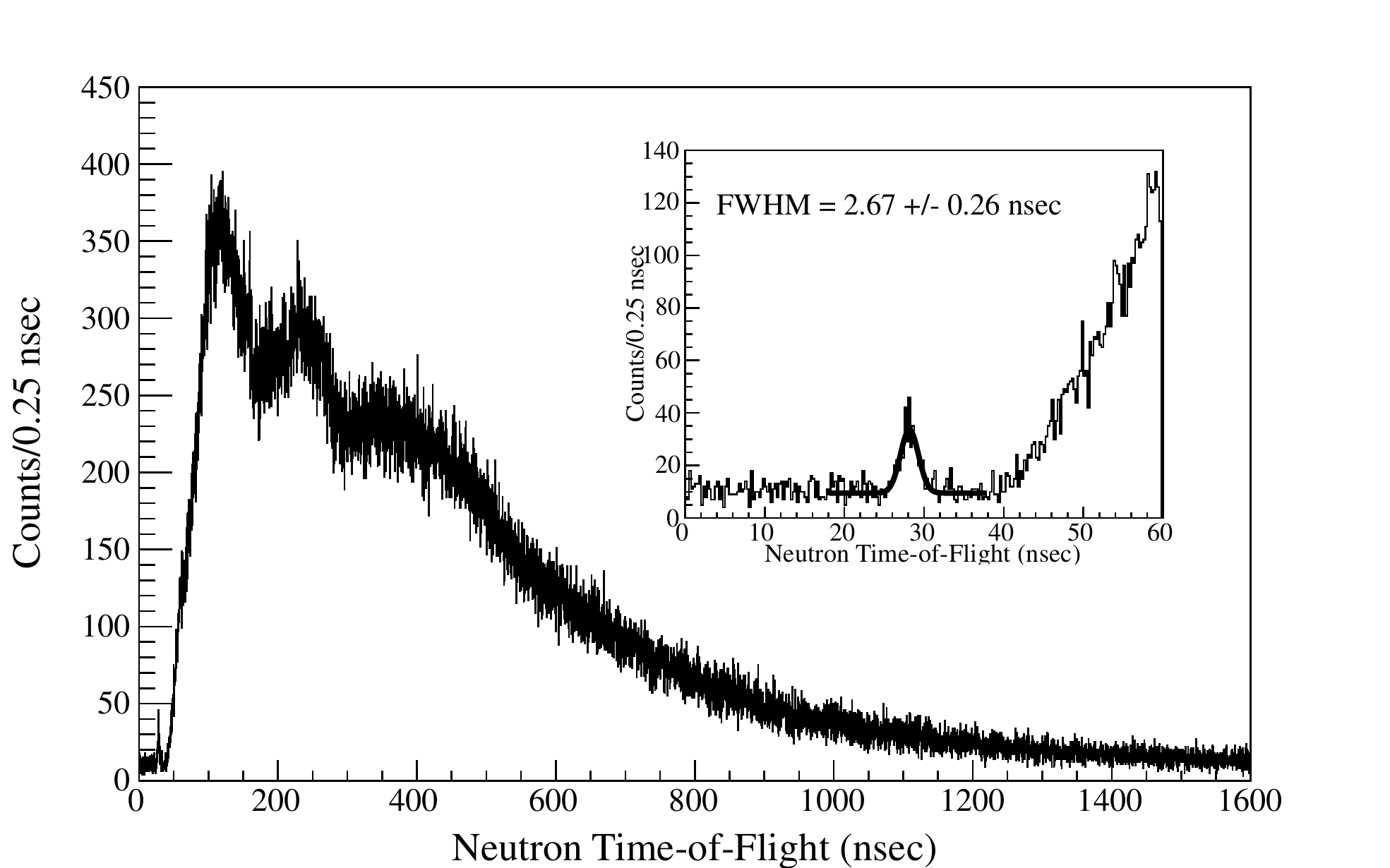}
    \caption{The time of flight distribution from the fissionTPC.  
    The inset shows the photo fission peak and the extracted timing resolution.
    }
    \label{fig:tof}
  \end{center}
\end{figure}

The cathode signal from the modified Cremat preamplifier (section~\ref{electronics}) has a rise time of about 10~ns when tested with a square wave pulse generator and capacitor charge injector.  
The observed rise time when connected to the fissionTPC cathode in operation is about 80--100~ns from a fission fragment.  
The exact cause of this difference in rise time is under investigation, but one should note that this has not prevented time of flight measurements with a resolution on the order of 1~ns (Fig.~\ref{fig:tof}).

The TOF resolution is easily measured by looking at the width of the photo fission peak in the inset of Fig.~\ref{fig:tof} because the intrinsic width of the peak is 250~ps.  
The distance of the fissionTPC from the neutron production target is also important in the measurement of TOF and this is accurately determined by placing a carbon block in the beam before the fissionTPC which scatters neutrons from a very narrow, well-known energy at 2078.0$\pm$0.3~keV that is well known in energy.  
For one placement of the fissionTPC, the time difference from the photo fission peak to the carbon notch was measured to be 390.3$\pm$0.2~ns which corresponds to a distance from the source to the fissionTPC cathode of 8336$\pm$4~mm.

\section{Conclusions}

The fissionTPC has been built for high precision, high accuracy (sub-percent total uncertainty) cross section measurements of the major actinides.  
A number of challenging design constraints have been incorporated to accommodate high rate, high ionization events and the neutron beam environment.  
The fissionTPC has demonstrated stable operation in the neutron environment and the dynamic range to measure alpha particles and fission fragments.
The fissionTPC has made preliminary beam and target uniformity measurements as well as rudimentary particle identification; both of these are important to understanding the systematic uncertainties inherent in cross section measurements.
The performance of the fissionTPC thus far indicates that the proposed cross section measurement campaign will likely succeed, once the improvements discussed in the text are implemented.  

\subsection{Additional Measurements possible with the fissionTPC}

Once a significant investment has been made for an instrument such as the fissionTPC it is natural and cost effective to think of other measurements that could be made.  
Without modification and perhaps even with the data already collected, other measurements are possible: ternary fission, mass and charge distributions as a function of neutron energy, and fission fragment angular distributions.  
It is anticipated that the experimental program will also be extended to (n,f) measurements of other actinides.

\section{Acknowledgments}

The authors would like to thank the following for their efforts in making this project possible: 
Dennis McNabb, Ed Hartouni, and John Becker for the idea of using a TPC to measure fission cross sections and Peter Barnes for working on the first feasibility study.
This work was performed under the auspices of the U.S. Department of Energy by Lawrence Livermore National Laboratory in part under Contract W-7405-Eng-48 and in part under Contract DE-AC52-07NA27344.
This work has also benefited from the use of the Los Alamos Neutron Science Center at the Los Alamos National Laboratory. 
This facility is funded by the US Department of Energy and operated by Los Alamos National Security, LLC under contract DE-AC52-06NA25396.
The NIFFTE university groups were supported through the DOE NERI program and through subcontracts from LLNL, LANL, and INL.
INL and PNNL were funded in part by the U.S. Department of Energy, Office of Nuclear Energy. LLNL-JRNL-651187





\bibliographystyle{elsarticle-num.bst}
\bibliography{fissionTPC.bib}






\end{document}